\documentclass[12pt]{article}
\usepackage[a4paper, total={7in, 10in}]{geometry}
\usepackage[parfill]{parskip}
\usepackage{physics, tensor, float, subcaption}
\usepackage{graphicx}
\graphicspath{ {Plots/} }
\usepackage{jhep-mod}
\usepackage{bm}
\usepackage{soul}
\usepackage{mathrsfs}
\usepackage[utf8]{inputenc}
\usepackage{enumerate}
\usepackage{bigints}
\usepackage{xcolor}
\usepackage{appendix}
\usepackage{graphicx}
\usepackage{float}
\usepackage{tikz}
\usepackage{setspace}
\usepackage{cancel}
\usepackage{doi}
\usepackage{amssymb,amsmath,amsthm}
\definecolor{purple}{rgb}{1,0,1}
\definecolor{lime}{HTML}{A6CE39} 

\newcommand{\blue}[1]{{\color{blue} #1}}

\definecolor{lime}{HTML}{A6CE39}
\newcommand{\orcidicon}{%
	\begin{tikzpicture}
	\draw[lime, fill=lime] (0,0) 
		circle [radius=0.16] 
		node[white] {{\fontfamily{qag}\selectfont \tiny ID}};
	\draw[white, fill=white] (-0.0625,0.095) 
		circle [radius=0.007];
	\end{tikzpicture}
	\hspace{-5mm}
}
\newcommand\orcidMatt{{\href{https://orcid.org/0000-0003-1088-6485}{\orcidicon}}}

\begin{document}

\title{\null\vspace{-50pt}
\leftline{\blue{Explicit formulae for stochastic equilibria}}}

\author{
\Large
Matt Visser\!\orcidMatt\!
}
\affiliation{School of Mathematics and Statistics, Victoria University of Wellington, \\
\null\qquad PO Box 600, Wellington 6140, New Zealand.}
\emailAdd{matt.visser@sms.vuw.ac.nz}
\def\theta{\vartheta}
\def\cof{{\mathrm{cofactor}}}
\def\adj{{\mathrm{adjugate}}}
\def\rk{{\mathrm{rank}}}

\abstract{
\vspace{1em}

Finding the stochastic equilibria for finite-state stochastic matrices amounts to solving an eigen\-vector problem $\pi = \pi P$.
Various techniques for doing so are known, some extremely computationally intensive. 
Herein we shall aim to extract a number of relatively simple analytic results that shed light on this problem. 
It is very easy to find an explicit general formula for the equilibrium vector (when it is unique) of a $2\times 2$ stochastic matrix. 
The corresponding explicit general formula for the equilibrium vector (when it is unique) of a $3\times 3$ stochastic matrix is a somewhat messier four-line result. (Though with a bit of work you can shoe-horn it into one line of text.)
An explicit general formula for the equilibrium vector (when it is unique) of a $4\times 4$ stochastic matrix requires a paragraph of text.
Ultimately, for $n\times n$ stochastic matrices a general and fully explicit construction of the equilibrium vector (when it is unique) can be developed in terms of a suitable adjugate (classical adjoint) matrix, and can subsequently be reduced to the computation of $n$ principal matrix minors.
Finally, an application to random walks on graphs is presented.

\bigskip

\bigskip
\noindent
{\sc Date:} Tuesday 27 January 2026; \LaTeX-ed \today


}

\maketitle
\def\tr{{\mathrm{tr}}}
\def\diag{{\mathrm{diag}}}
\def\pdet{{\mathrm{pdet}}}
\def\d{{\mathrm{d}}}
\parindent0pt
\parskip7pt

\def\Z{{\mathbb{Z}}}

\clearpage
\section{Introduction}
The study of stochastic matrices (Markov matrices) has a long and rather complex history, within both mathematical physics and pure mathematics~\cite{Stochastic,Stone,Ayers,Anton,Leon,Z-matrix,Topics,Analysis,Functions,Adjugate}. 
Finding the stochastic equilibria for finite-state systems amounts to solving an eigen\-vector problem $\pi = \pi P$.
Here we take $P$ to be a (row) stochastic matrix, with non-negative entries and each row summing to 1. Defining $u=[1,\dots,1]^T$, the normalization condition $Pu=u$ implies that 1 is an eigenvalue of the stochastic matrix $P$, and $u$ is the corresponding  right eigenvector. We are interested in evaluating the corresponding left eigenvector $\pi$. 

Various ways of proceeding are well-known. Popular methods involve brute force extraction of eigenvectors~\cite{Stochastic,Stone,Ayers,Anton,Leon,Functions,Z-matrix,Topics,Analysis,Adjugate}, and the so-called ``power method'' where one considers the sequence
\begin{equation}
P \to P^2 \to (P^2)^2= P^4 \to (P^4)^2 = P^8 \to \dots \to P^{2^m}\to \cdots...
\end{equation}
If the one explicitly known eigenvalue 1 has multiplicity 1, and all other eigenvalues satisfy the strict inequality $|\lambda|<1$, then $\lim_{m\to\infty}  P^{2^m} = u \pi$. This naive application of the power method fails, for instance, for permutation matrices, and gives unexpected answers when the eigenvalue 1 has multiplicity greater than 1, (for instance, when $P$ is block diagonalizable by permutations). In numerical applications one often forces the issue by adding an ``infinitesimal'' perturbation to the stochastic matrix $P$ to guarantee the sub-dominant eigenvalues satisfy $|\lambda|<1$. 

Below I shall, to the extent possible, develop direct algorithmic techniques that do not involve taking limits. 
Dealing with $2\times 2$ stochastic matrices is extremely easy, and the results are well-known. 
We shall then present a similar (slightly more complicated but still very tractable) explicit formula for the equilibrium vector of 
$3\times 3$ stochastic matrices. Dealing with $4\times 4$ stochastic matrices is relatively messy, and the explicit formula for $5\times 5$ stochastic matrices is quite awful.  
Nevertheless useful things can be said for $n\times n$ stochastic matrices --- we shall present a general construction of the equilibrium vector (when it is unique) in terms of a suitable adjugate (classical adjoint) matrix~\cite{Stochastic,Stone,Ayers,Anton,Topics,Analysis,Adjugate}.
This construction can subsequently be reduced to the computation of the $n$ principal minors of a suitable matrix.

We shall then consider random walks on graphs, with the stochastic matrix being defined in terms of the adjacency matrix and degree matrix --- certain parts of the analysis simplify. (The computation can be reduced to dealing with integer matrices and their principal minors.) 

\bigskip
\hrule\hrule\hrule

\clearpage
\section{Stochastic equilibria for $2\times 2$ stochastic matrices}

Any $2\times 2$ (row) stochastic matrix can be written in the form
\begin{equation}
P_2 = \left[\begin{array}{cc}1-p& p\\ q &1-q\end{array}\right]
\qquad
0 \leq \{p,q\} \leq 1.
\end{equation}
It is then easy to see (and is reasonably well-known) that the (row) vector
\begin{equation}
\label{E:pi2}
\pi_2 = {\left[ q, \; p \right]\over p+q},
\end{equation}
when it exists, is the unique normalized equilibrium probability vector satisfying
\begin{equation}
\pi_2 \; P_2 = \pi_2. 
\end{equation}
The only way this construction can fail is if the denominator is zero: $p+q=0$. 
This exceptional case corresponds to $P_2 \to I$, for which the equilibrium is not unique. This degeneracy can always be lifted (if desired) by an infinitesimal perturbation. For instance, consider
$(p,q)=(0,0)\to (\epsilon,\epsilon)$  with $ 0 < \epsilon \ll 1$. Then $\pi_2 \to [\frac12,\frac12]$.
More generally for $P_2\to I$ \emph{any} probability 2-vector is an equilibrium.
Variants of this infinitesimal perturbation construction for lifting possible equilibrium degeneracy can also be developed for $n\times n$ stochastic matrices. 

In summary, as long as a unique equilibrium for $2\times2$ stochastic matrices exists, it is given by equation (\ref{E:pi2}). 
If the numerator of equation (\ref{E:pi2}) is zero, this forces $P_2\to I_2$  and is a diagnostic for the existence of multiple equilibria. These observations generalize to higher dimensionality.

\bigskip
\hrule\hrule\hrule

\clearpage
\section{Stochastic equilibria for $3\times 3$ stochastic matrices}
\subsection{General situation}
Any $3\times 3$ (row) stochastic matrix can be written in the form
\begin{equation}
P_3 = \left[\begin{array}{ccc}1-p_1-p_2& p_1& p_2\\ q_2 &1-q_1-q_2& q_1\\ r_1 & r_2 & 1-r_1- r_2\end{array}\right]
\end{equation}
with $0 \leq \{p_i,q_i,r_i\} \leq 1$ and $\{p_1+p_2,q_1+q_2,r_1+r_2\} \leq 1$.

Now define three non-negative quantities
\begin{equation}
w_1 = q_1 r_1 + q_2 r_1 + q_2 r_2;
\end{equation}
\begin{equation}
w_2 = r_1 p_1 + r_2 p_1 + r_2 p_2;
\end{equation}
\begin{equation}
w_3 = p_1 q_1 + p_2 q_1 + p_2 q_2;
\end{equation}
and define the row vector
\begin{equation}
\label{E:pi3}
\pi_3 = {[w_1,w_2,w_3]\over w_1+w_2+w_3}.
\end{equation}
Then as long as $w_1+w_2+w_3$ is nonzero, $\pi_3$ is a properly normalized (row) probability vector and it is easy (if slightly tedious) to check that it is the equilibrium probability vector
\begin{equation}
\pi_3 \; P_3 = \pi_3. 
\end{equation}

It is useful to note the symmetry
\begin{equation}
\label{E:permute3}
w_1\leftrightarrow w_2\leftrightarrow w_3: \qquad (q_i,r_i) \leftrightarrow (r_i,p_i) \leftrightarrow (p_i,q_i).
\end{equation}
All in all, this is a remarkably compact expression for three dimensional stochastic equilibria.
Explicitly, we have the one-line result:
\begin{equation}
\label{E:pi3b}
\pi_3 = {[q_1 r_1 + q_2 r_1 + q_2 r_2, \;\;\; r_1 p_1 + r_2 p_1 + r_2 p_2, \;\;\;p_1 q_1 + p_2 q_1 + p_2 q_2]
\over 
(q_1 r_1 + q_2 r_1 + q_2 r_2)+( r_1 p_1 + r_2 p_1 + r_2 p_2)+(p_1 q_1 + p_2 q_1 + p_2 q_2)}. 
\end{equation}
This expression was first found by brute force evaluation and inspection of the eigenvectors of $P_3$.
Below we shall develop an understanding of where this result is coming from, and generalize it in various ways. 

\clearpage
\subsection{Exceptional cases}

\leftline{Let us now consider the exceptional cases where one or more of the $w_i$ is/are zero.}

First, let us start with one of the $w_i$ being zero, and in view of the symmetry (\ref{E:permute3}) above permute rows and columns so that it is $w_1$ that is zero. We then have
\begin{equation}
w_1=0 = q_1 r_1 + q_2 r_1 + q_2 r_2 = (q_1 + q_2) r_1 + q_2 r_2 = q_1 r_1 + q_2 (r_1+ r_2).
\end{equation}
There are only a limited number of ways this can be achieved.
\begin{itemize}
\item $q_1=q_2=0$: This would imply $w_3=0$, while $w_2$ is unconstrained.\\
The equilibrium vector (provided $w_2>0$) would then reduce to $\pi=[0,1,0]$.\\
The probability matrix reduces to
\begin{equation}
P_3 = \left[\begin{array}{ccc}1-p_1-p_2\;\;\;& p_1& p_2\\ 0 &1& 0\\ r_1 & r_2 & \;\;\;1-r_1- r_2\end{array}\right],
\end{equation}
so that ``2'' is an absorbing state.
\item $r_1=r_2=0$: This would imply $w_2=0$, while $w_3$ is unconstrained.\\
The equilibrium vector (provided $w_3>0$) would then reduce to $\pi=[0,0,1]$.\\
The probability matrix reduces to
\begin{equation}
P_3 = \left[\begin{array}{ccc}1-p_1-p_2& p_1& p_2\\ q_2 &1-q_1-q_2\;\;\;& q_1\\ 0 & 0 & 1\end{array}\right],
\end{equation}
so that ``3'' is an absorbing state.
\item $r_1=q_2=0$: This would imply $w_2= r_2(p_1+p_2)$ and $w_3=q_1(p_1+p_2)$.\\
The equilibrium vector would reduce to $\pi=[0,r_2,q_1]/(r_2+q_1)$.\\
The probability matrix reduces to
\begin{equation}
P_3 = \left[\begin{array}{ccc}1-p_1-p_2& p_1& p_2\\ 0 &\;\;\;1-q_1\;\;\;& q_1\\ 0 & r_2 & 1- r_2\end{array}\right],
\end{equation}
so that (if $p_1+p_2>0$) the ``2''--``3'' sub-matrix is absorbing. Note that the first row is now transitory and that the equilibrium vector is simply that of the ``2''--``3'' sub-matrix.  Explicitly $\pi=[0,r_2,q_1]/(q_1+r_2)$,
\end{itemize}

\enlargethispage{10pt}
Second, now consider the case where all three of the $w_i$ are zero. We have already seen how to make two of the $w_i$ zero, so without loss of generality let us permute states to make $w_1=w_2=0$ (corresponding to $r_1=r_2=0$) and ask what more would be needed to make $w_3=0$. 

This would additionally require
\begin{equation}
w_3 = 0= p_1 q_1 + p_2 q_1 + p_2 q_2 = (p_1 +p_2) q_1 + p_2 q_2 = p_1 q_1 + p_2 (q_1 + q_2). 
\end{equation}
There are only a limited number of ways this can be achieved.
\begin{itemize}
\item $p_1=p_2=0$: In this situation the probability matrix reduces to
\begin{equation}
\label{E:P3-exception1}
P_3 = \left[\begin{array}{ccc}1& 0& 0\\ q_2 &\;\;\;1-q_1-q_2\;\;\;& q_1\\ 0 & 0 & 1\end{array}\right].
\end{equation}
Note that there are two distinct absorbing states, ``1'' and ``3'', and so two distinct equilibria $[1,0,0]$ and $[0,0,1]$.  No unique equilibrium exists.

\item $q_1=q_2=0$: In this situation the probability matrix reduces to
\begin{equation}
\label{E:P3-exception2}
P_3 = \left[\begin{array}{ccc}1-p_1-p_2\;\;\;& p_1& p_2\\ 0&1&0 \\ 0 & 0 & 1\end{array}\right].
\end{equation}
Note that there are two distinct absorbing states, ``2'' and ``3'', and so two distinct equilibria $[0,1,0]$ and $[0,0,1]$.  No unique equilibrium exists.

\item $q_1=p_2=0$: In this situation the probability matrix reduces to
\begin{equation}
\label{E:P3-exception3}
P_3 = \left[\begin{array}{ccc}1-p_1& p_1& 0\\ q_2 &1-q_2\;\;\;& 0\\ 0 & 0 & 1\end{array}\right].
\end{equation}
This is a direct sum of a $2\times2$ stochastic block in the ``1''--``2'' subspace, and a singleton absorbing state ``3''. There are now two distinct equilibria. Explicitly, $[q_2,p_1,0]/(p_1+q_2)$ and $[0,0,1]$. 
 No unique equilibrium exists.
 If we now additionally assume $p_1=q_2=0$ (which is more than is needed to make all three $w_i$ zero), then $P_3\to I_3$, and there are three distinct equilibria $[1,0,0]$, $[0,1,0]$, and $[0,0,1]$. 
\end{itemize}

\subsection{Summary}

\enlargethispage{35pt}
As long as a unique equilibrium for $3\times3$ stochastic matrices exists, it is given by equation (\ref{E:pi3}), or the fully explicit version of equation (\ref{E:pi3b}). This corresponds to the stochastic matrix being irreducible.
(Either $P_3>0$ is strictly positine, or for some finite integer power $(P_3)^k > 0$ is strictly positive.)
If the numerator of equation (\ref{E:pi3}) or  equation (\ref{E:pi3b}) is zero, this is a diagnostic for the existence of multiple equilibria; the stochastic matrix is then reducible.
Let us now extend the discussion, first to $4\times4$ stochastic matrices, and then to the general $n\times n$ case, where we shall see why this is happening so nicely.

\bigskip
\hrule\hrule\hrule

\section{Stochastic equilibria for $4\times 4$ stochastic matrices}

Any $4\times 4$ (row) stochastic matrix can be written in the form
\begin{equation}
P = \left[\begin{array}{cccc}1-p_1-p_2-p_3& p_1& p_2&p_3\\ q_3 &1-q_1-q_2-q_3& q_1&q_2\\ 
r_2 & r_3 & 1-r_1- r_2-r_3&r_1\\ s_1&s_2&s_2 &1-s_1-s_2-s_3\end{array}\right]
\end{equation}
\leftline{with $0 \leq \{p_i,q_i,r_i,s_i\} \leq 1$ and $\{p_1+p_2+p_3,q_1+q_2+q_3,r_1+r_2+r_3,s_1+s_2+s_3\} \leq 1$.}

Now define four non-negative quantities
\begin{eqnarray}
w_1 &=& q_1  r_1  s_1 +  q_1  r_2  s_1 +  q_1  r_2  s_2 +  q_1  r_2  s_3 +  q_2  r_1  s_1 +  q_2  r_2  s_1 +  q_2  r_2  s_3 +  q_2  r_3  s_1 \nonumber\\
&+&  q_3  r_1  s_1 +  q_3  r_1  s_2 +  q_3  r_2  s_1 +  q_3  r_2  s_2 +  q_3  r_2  s_3 +  q_3  r_3  s_1 +  q_3  r_3  s_2 +  q_3  r_3  s_3;\qquad
\end{eqnarray}
\begin{eqnarray}
w_2 &=&  p_1  r_1  s_1 +  p_1  r_1  s_2 +  p_1  r_2  s_1 +  p_1  r_2  s_2 +  p_1  r_2  s_3 +  p_1  r_3  s_1 +  p_1  r_3  s_2 +  p_1  r_3  s_3 \nonumber\\
&+&  p_2  r_1  s_2 +  p_2  r_3  s_1 +  p_2  r_3  s_2 +  p_2  r_3  s_3 +  p_3  r_1  s_2 +  p_3  r_2  s_2 +  p_3  r_3  s_2 +  p_3  r_3  s_3; \qquad
\end{eqnarray}
\begin{eqnarray}
w_3 &=&  p_1  q_1  s_1 +  p_1  q_1  s_2 +  p_1  q_1  s_3 +  p_1  q_2  s_3 +  p_2  q_1  s_1 +  p_2  q_1  s_2 +  p_2  q_1  s_3 +  p_2  q_2  s_1 \nonumber\\
&+&  p_2  q_2  s_3 +  p_2  q_3  s_1 +  p_2  q_3  s_2 +  p_2  q_3  s_3 +  p_3  q_1  s_2 +  p_3  q_1  s_3 +  p_3  q_2  s_3 +  p_3  q_3  s_3;\qquad
\end{eqnarray}
\begin{eqnarray}
w_4 &=&  p_1  q_1  r_1 +  p_1  q_2  r_1 +  p_1  q_2  r_2 +  p_1  q_2  r_3 +  p_2  q_1  r_1 +  p_2  q_2  r_1 +  p_2  q_2  r_3 +  p_2  q_3  r_1 \nonumber\\
&+&  p_3  q_1  r_1 +  p_3  q_1  r_2 +  p_3  q_2  r_1 +  p_3  q_2  r_2 +  p_3  q_2  r_3 +  p_3  q_3  r_1 +  p_3  q_3  r_2 +  p_3  q_3  r_3;\qquad
\end{eqnarray}
and define the row vector
\begin{equation}
\pi_4 = {[w_1,w_2,w_3,w_4]\over w_1+w_2+w_3+w_4}.
\end{equation}
Then as long as $w_1+w_2+w_3+w_4$ is nonzero, $\pi_4$ is a properly normalized (row) probability vector and it is easy (if now more than somewhat tedious) to check that it is the equilibrium probability vector
\begin{equation}
\label{E:pi4}
\pi_4 \; P_4 = \pi_4. 
\end{equation}

\enlargethispage{20pt}
It is useful to note the symmetry
\begin{equation}
\label{E:permute4}
w_1\leftrightarrow w_2\leftrightarrow w_3\leftrightarrow w_4: \qquad 
(q_i,r_i,s_i) \leftrightarrow (r_i,s_i,p_i) \leftrightarrow (s_i,p_i,q_i)\leftrightarrow (p_i,q_i,r_i).
\end{equation}
Each of the $w_i$ now contains 16 terms, so the denominator $\sum w_i$ contains 64 terms.

These explicit formulae are now sufficiently complex so as to be mildly unwieldy. Their existence is more useful as a ``proof of principle'', and a guide towards the general analysis to be developed below.

Even worse, if one were to consider $5\times 5$ stochastic matrices, then there would be five $w_i$, and we shall soon see that each of these $w_i$ would contain 125 terms, so the denominator $\sum w_i$ would contain 625 terms.
It is clear that a more general analysis is required. 

\bigskip
\hrule\hrule\hrule

\section{Stochastic equilibria for $n\times n$ stochastic matrices}
\subsection{Generic situation}

For a generic $n\times n$ stochastic matrix we have
\begin{equation}
P_n = \left[\begin{array}{ccccc}
p_{11}& p_{12}&\cdots &p_{1(n-1)} & p_{1n}\\
p_{21}& p_{22} &\cdots &p_{2(n-1)} & p_{2n}\\
\vdots & \vdots &\ddots & \vdots&\vdots\\
p_{(n-1)1} &p_{(n-1)2}&\cdots & p_{(n-1)(n-1)}& p_{(n-1)n}\\
p_{n1} &p_{n2}&\cdots & p_{n(n-1)}&p_{nn}

\end{array}\right];
\qquad 
u_n = \left[\begin{array}{c}
1\\
1\\
\vdots\\
1\\
1
\end{array}\right];
\end{equation}
with $p_{ij}\geq 0$ and $P_n u_n = u_n$, that is $\sum_j p_{ij}=1$. 

It is useful to eliminate the diagonal entries by using $p_{ii} = 1 - \sum_{j\neq i} p_{ij}$ to write
\begin{equation}
P_n = \left[\begin{array}{ccccc}
1 - \sum_{j\neq 1} p_{1j}&p_{12} &\cdots &p_{1(n-1)}& p_{1n}\\
p_{21} &1 - \sum_{j\neq 1} p_{2j} &\cdots &p_{2(n-1)}& p_{2n}\\
\vdots & \vdots &\ddots & \vdots&\vdots\\
p_{(n-1)1} &p_{(n-1)2} &\cdots & 1 - \sum_{j\neq n-1} p_{(n-1)j}& p_{(n-1)n}\\
p_{n1} &p_{n2}&\cdots & p_{n(n-1)} & 1 - \sum_{j\neq n} p_{nj}
\end{array}\right].
\end{equation}
Doing this parameterises the stochastic matrix $P_n$ in terms of the $p_{i\neq j}$, 
subject to the constraints $p_{i\neq j}\geq 0$ and 
$ \sum_{j\neq i} p_{ij} \leq 1$. 

Now note that
\begin{equation}
I_n - P_n = \left[\begin{array}{ccccc}
\;\sum_{j\neq 1} p_{1j}\;&-p_{12} &\cdots &-p_{1(n-1)}& -p_{1n}\\
-p_{21} &\;\sum_{j\neq 1} p_{2j} \;&\cdots &-p_{2(n-1)}& -p_{2n}\\
\vdots & \vdots &\;\ddots\; & \vdots&\vdots\\
-p_{(n-1)1} &-p_{(n-1)2} &\cdots & \;\sum_{j\neq n-1} p_{(n-1)j}\;& -p_{(n-1)n}\\
-p_{n1} &-p_{n2}&\cdots &- p_{n(n-1)} & \;\sum_{j\neq n} p_{nj} \;
\end{array}\right].
\end{equation}

\clearpage
This is a specific example of a so-called Z-matrix (non-negative diagonal, non-positive off-diagonal)~\cite{Topics,Z-matrix}.
Quite a lot is known regarding the behaviour of Z-matrices, and we shall use key results as needed below~\cite{Topics,Z-matrix}.

Now is a good time to remind the reader of the definition of the adjugate matrix (the transpose of the cofactor matrix, sometimes called the classical adjoint)~\cite{Stone,Ayers,Analysis,Adjugate}. For any arbitrary $n\times n$ matrix $X_n$, in terms of the matrix minors $M_{ij}(X_n)$ we have
\begin{equation}
\cof(X_n) = [ (-1)^{i+j} M_{ij}(X_n) ]; \qquad \adj(X_n) = [ (-1)^{i+j} M_{ji}(X_n) ].
\end{equation}
The key result that we will use and re-use multiple times is that~\cite{Stone,Ayers,Analysis,Adjugate}. 
\begin{equation}
\adj(X_n) \;X_n = X_n \;\adj(X_n) = \det(X_n) \; I_n.
\end{equation}

Note that by construction $(I_n-P_n) u_n=0$, so $I_n-P_n$ is a singular matrix. Thence
\begin{equation}
\adj(I_n-P_n) \;(I_n-P_n) = (I_n-P_n) \;\adj(I_n-P_n) = 0.
\end{equation}
Provided the equilibrium vector is unique we know that the dominant eigenvalue of $P_n$, namely 1, has multiplicity 1.
This implies $\rk(I_n-P_n)=n-1$, which in turn implies $\rk(\adj(I_n-P_n))=1$.\\
\emph{Proof:} By construction the left eigenvectors of $\adj(I_n-P_n)$ are orthogonal to the right eigenvectors of $I_n-P_n$,
and the right eigenvectors of $\adj(I_n-P_n)$ are orthogonal to the left eigenvectors of $I_n-P_n$. But note that the eigenvectors of $I_n-P_n$ span a $n-1$ dimensional vector space, so the eigenvectors of the adjugate matrix $\adj(I_n-P_n)$, being orthogonal thereto, can at most span a $1$ dimensional vector space. 
\hfill{$\Box$}

But we know that $u_n$ and $\pi_n$ are eigenvectors of $I_n-P_n$ with eigenvalue 0. Thence we must have
\begin{equation}
\adj(I_n-P_n) \propto u_n \pi_n. 
\end{equation}
Normalizing, we see
\begin{equation}
  u_n \pi_n = {\adj(I_n-P_n) \over \tr(\adj(I_n-P_n))}.
\end{equation}
But then, working in terms of the principal minors of $I_n-P_n$ we have the explicit result
\begin{equation}
  \pi_n = {[M_{1,1}(I_n-P_n), \cdots, M_{ii}(I_n-P_n), \cdots, M_{nn}(I_n-P_n) ] \over \sum_i M_{ii}(I_n-P_n)}.
\end{equation}

\clearpage
This is of the desired form
\begin{equation}
  \pi_n = {[w_1, \cdots, w_{i}, \cdots, w_{n}] \over \sum_i w_{i}},
\end{equation}
where explicitly the weights $w_i$ are just the principal matrix minors:
\begin{equation}
  w_i = M_{ii}(I_n-P_n).
\end{equation}
Note that, since $I_n-P_n$ is a $n\times n$ Z-matrix, deleting the $ith$ row and $ith$ column yields a $(n-1)\times(n-1)$ Z-matrix whose row sums are non-negative. Therefore its determinant is non-negative --- all of the principal minors of $I_n-P_n$ are non-negative --- and so $\pi_n$ is a normalized non-negative vector as required~\cite{Topics}. 

In summary, to evaluate the stochastic equilibrium of an $n\times n$ stochastic matrix $P_n$, all you need to do is to calculate the $n$ principal minors of $I_n-P_n$. This might be a little tedious, but is utterly straightforward. 
In fact, if all you need is the \emph{relative} equilibrium probabilities of two states, then you only need to calculate the ratio of two principal minors:
\begin{equation}
{(\pi_n)_i\over(\pi_n)_j} = {M_{ii}(I_n-P_n)\over M_{jj}(I_n-P_n)}.
\end{equation}

\subsection{Exceptional cases}

The only situation where the algorithm fails is when \emph{all} of the principal minors $M_{ii}(I_n-P_n)$ are zero, in which case $\pi = [0,\cdots, 0]/0$ is undefined. 
But if all of the $M_{ii}(I_n-P_n)=0$, this means that 0 must be an eigenvalue of each principal sub-matrix of $(I_n-P_n)$.
So 0 must be a repeated eigenvalue of $(I_n-P_n)$ itself, so 1 must be a repeated eigenvalue of $P_n$. 
Note that this implies the adjugate matrix of $(I_n-P_n)$ must be identically zero:  $\adj(I_n-P_n)=0$. 

\enlargethispage{15pt}
The simplest situation where this happens is if the stochastic matrix can be permuted into block diagonal form:
\begin{equation}
P = \left[\begin{array}{c|c}P_a&0\\ \hline 0 & P_b \end{array}\right].
\end{equation}
If $P_a$ and $P_b$ are themselves irreducible they will individually have unique equilibria $\pi_a = \pi_a P_a$ and $\pi_b = \pi_b P_b$. Then $P$ has two distinct specific equilibria, namely $[\pi_a|0]$ and $[0|\pi_b]$, which can be used to span the entire set of all equilibria:
\begin{equation}
\pi(p) = [p\pi_a|(1-p)\pi_b], \qquad p\in[0,1], \qquad \hbox{with} \qquad \pi(p) = \pi(p) P.
\end{equation}
If either $P_a$ or $P_b$ further block diagonalize, then the process continues. The most exceptional  situation that can happen is $P\to I_n$ in which case any probability vector is an equilibrium.

\clearpage
A slightly more subtle way of getting multiple equilibria is in the presence of transitory states.
Suppose we can permute the stochastic matrix into the form
\begin{equation}
P = \left[\begin{array}{c|c|c}
T_a&T_b&T_c\\
\hline
0&P_a&0\\ 
\hline 0 &0 & P_b 
\end{array}\right].
\end{equation}
Then if $T_a$, $T_b$, $T_c$ are strictly positive sub-matrices the corresponding states will be transitory, 
and the equilibria are determined purely in terms of the $P_a\oplus P_b$ sub-matrix. 
Then $P$ has two distinct specific equilibria, namely $[0|\pi_a|0]$ and $[0|0|\pi_b]$, which can be used to span the set of all equilibria:
\begin{equation}
\pi(p) = [0|p\pi_a|(1-p)\pi_b], \qquad p\in[0,1], \qquad \hbox{with} \qquad \pi(p) = \pi(p) P.
\end{equation}

We have already seen specific examples of both types of exceptional behaviour when analyzing $2\times 2$ and $3\times 3$ stochastic matrices. See equations (\ref{E:pi2}), (\ref{E:P3-exception1}), (\ref{E:P3-exception2}), (\ref{E:P3-exception3}), and related discussion. 

Overall, we have rather tight control over where exceptional cases occur, and good strategies for dealing with these issues --- just find all of the irreducible sub-matrices. Once you have identified all the irreducible sub-matrices, each one of them will have a unique equilibrium vector which can be explicitly calculated by the algorithm above. Embed these  equilibrium vectors in $n$ dimensions in the canonical way, and take the convex polytope thereof. This will characterize the general equilibria of the original reducible stochastic matrix $P$.

\subsection{Some straightforward consistency checks}

\begin{itemize}
\item For $n=2$  we have
\begin{equation}
I_2-P_2 = \left[\begin{array}{cc}p&-p\\-q&q\end{array}\right]; \qquad 
w_1 = M_{11}(I_2-P_2)=q; \quad w_2 = M_{22}(I_2-P_2)=p.
\end{equation}
This reproduces the result presented in equation (\ref{E:pi2}).

\item For $n=3$ we have
\begin{equation}
I_3-P_3 = \left[\begin{array}{ccc}p_1+p_2& -p_1& -p_2\\ -q_2 &q_1+q_2& -q_1\\ -r_1 & -r_2 & r_1+ r_2\end{array}\right]
\end{equation}
Then (as required)
\begin{equation}
w_1 = M_{11}(I_3-P_3) = \det \left[\begin{array}{ccc} q_1+q_2& -q_1\\  -r_2 & r_1+ r_2\end{array}\right]
= q_1 r_1 + q_2 r_1 + q_2 r_2 .
\end{equation}
Then $w_2$ and $w_3$ follow by the permutation symmetry in equation (\ref{E:permute3}).

\item For $n=4$ we have
\begin{equation}
I_4-P_4 = \left[\begin{array}{cccc}
p_1+p_2+p_3& -p_1& -p_2& -p_3\\ 
-q_3 &q_1+q_2+q_3& -q_1&-q_2\\
 -r_2 & -r_3  & r_1+ r_2+r_3 &-r_1\\
 -s_1& -s_2 & -s_3  & s_1+ s_2+s_3 
 \end{array}\right].
\end{equation}
Then
\begin{equation}
w_1 = M_{11}(I_4-P_4) = \det \left[\begin{array}{ccc}
q_1+q_2+q_3& -q_1&-q_2\\
-r_3  & r_1+ r_2+r_3 &-r_1\\
-s_2 & -s_3  & s_1+ s_2+s_3 
 \end{array}\right].
\end{equation}
This now easily (if tediously) evaluates (as required) to 
\begin{eqnarray}
w_1  &=& M_{11}(I_4-P_4) \\
&=& q_1  r_1  s_1 +  q_1  r_2  s_1 +  q_1  r_2  s_2 +  q_1  r_2  s_3 +  q_2  r_1  s_1 +  q_2  r_2  s_1 +  q_2  r_2  s_3 +  q_2  r_3  s_1\qquad \nonumber\\
&+&  q_3  r_1  s_1 +  q_3  r_1  s_2 +  q_3  r_2  s_1 +  q_3  r_2  s_2 +  q_3  r_2  s_3 +  q_3  r_3  s_1 +  q_3  r_3  s_2 +  q_3  r_3  s_3. \nonumber
\end{eqnarray}
Then $w_2$, $w_3$ and $w_4$ follow by the permutation symmetry in equation (\ref{E:permute4}). 
\item
I had previously alluded to what happens for $5\times5$ stochastic matrices. Here are the details. 
Any $5\times 5$ (row) stochastic matrix can be written in the form
\begin{equation}
P_5 = \left[\begin{array}{ccccc}
\;1-\sum p_i\;& p_1& p_2&p_3&p_4\\ 
q_4 &\;1-\sum q_i\;& q_1&q_2&q_3\\ 
r_3 & r_4 & \;1-\sum r_i \;&r_1&r_2\\ 
s_2&s_3&s_4 &\;1-\sum s_i \;& s_1\\
t_1&t_2&t_3 &t_4 &\;1-\sum t_i\;\end{array}\right],
\end{equation}
\leftline{with $0 \leq \{p_i,q_i,r_i,s_i,t_i\} \leq 1$ and $\{\sum p_i, \sum q_i,\sum r_i ,\sum s_i, \sum t_i\} \leq 1$.}
Note the stochastic matrix $P_5$ exhibits permutation symmetry under all of the interchanges
{\small
\begin{equation}
(p_i,q_i,r_i,s_i,t_i) \leftrightarrow (q_i,r_i,s_i,t_i,p_i) \leftrightarrow (r_i,s_i,t_i,p_i,q_i)\leftrightarrow (s_i,t_i,p_i,q_i,r_i)
\leftrightarrow (t_i,p_i,q_i,r_i,s_i).
\end{equation} 
}
Note
\begin{equation}
I_5-P_5 = \left[\begin{array}{ccccc}
\;\sum p_i\;& -p_1&- p_2&-p_3&-p_4\\ 
-q_4 &\;\sum q_i\;& -q_1&-q_2&-q_3\\ 
-r_3 & -r_4 &\; \sum r_i\; &-r_1&-r_2\\ 
-s_2&-s_3&-s_4 &\;\sum s_i\; & -s_1\\
-t_1&-t_2&-t_3 &-t_4 &\;\sum t_i\;\end{array}\right]
\end{equation}
is a $5\times 5$ Z-matrix~\cite{Topics,Z-matrix}. Now define five quantities
\begin{equation}
w_1 = M_{11}(I_5-P_5) = \det\left[\begin{array}{ccccc}
\;\sum q_i\;& -q_1&-q_2&-q_3\\ 
 -r_4 & \;\sum r_i \;&-r_1&-r_2\\ 
-s_3&-s_4 &\;\sum s_i \;& -s_1\\
-t_2&-t_3 &-t_4 &\;\sum t_i\;\end{array}\right];
\end{equation}
\begin{equation}
w_2 = M_{22}(I_5-P_5) = \det \left[\begin{array}{ccccc}
\;\sum p_i\;& - p_2&-p_3&-p_4\\ 
-r_3 & \; \sum r_i\; &-r_1&-r_2\\ 
-s_2&-s_4 &\;\sum s_i\; & -s_1\\
-t_1&-t_3 &-t_4 &\;\sum t_i\;\end{array}\right];
\end{equation}
\begin{equation}
w_3 = M_{33}(I_5-P_5) = \det \left[\begin{array}{ccccc}
\;\sum p_i\;& -p_1&-p_3&-p_4\\ 
-q_4 &\;\sum q_i\;&-q_2&-q_3\\ 
-s_2&-s_3&\;\sum s_i\; & -s_1\\
-t_1&-t_2&-t_4 &\;\sum t_i\; \end{array}\right];
\end{equation}
\begin{equation}
w_4 = M_{44}(I_5-P_5) = \det \left[\begin{array}{ccccc}
\; \sum p_i\;& -p_1&- p_2&-p_4\\ 
-q_4 &\;\sum q_i\;& -q_1&-q_3\\ 
-r_3 & -r_4 & \;\sum r_i\; &-r_2\\ 
-t_1&-t_2&-t_3 &\;\sum t_i\;\end{array}\right];
\end{equation}
\begin{equation}
w_5 = M_{55}(I_5-P_5) = \det \left[\begin{array}{ccccc}
\;\sum p_i\;& -p_1&- p_2&-p_3\\ 
-q_4 &\;\sum q_i\;& -q_1&-q_2\\ 
-r_3 & -r_4 & \;\sum r_i \;&-r_1\\ 
-s_2&-s_3&-s_4 &\;\sum s_i\; \\
\end{array}\right].
\end{equation}
These principal matrix minors are all determinants of $4\times4$ Z-matrices (with non-negative row sums), and hence are guaranteed to be non-negative~\cite{Topics}.

It is useful to note the interchange symmetry $w_1\leftrightarrow w_2\leftrightarrow w_3\leftrightarrow w_4\leftrightarrow w_4$ 
\begin{equation}
\label{E:S5}
(q_i,r_i,s_i,t_i) \leftrightarrow (r_i,s_i,t_i,p_i) \leftrightarrow (s_i,t_i,p_i,q_i)\leftrightarrow (p_i,q_i,r_i,s_i).
\end{equation}
Explicitly, one can calculate:
{\tiny
\begin{eqnarray}
w_1 &=& 
q_1 r_1 s_1 t_1 + q_1 r_1 s_2 t_1 + q_1 r_1 s_2 t_2 + q_1 r_1 s_2 t_3 + q_1 r_1 s_2 t_4 + q_1 r_2 s_1 t_1 
+ q_1 r_2 s_2 t_1 + q_1 r_2 s_2 t_4 + q_1 r_2 s_3 t_1 + q_1 r_2 s_4 t_1 \nonumber\\
&+& 
q_1 r_3 s_1 t_1 + q_1 r_3 s_1 t_2 + q_1 r_3 s_1 t_3 + q_1 r_3 s_2 t_1 + q_1 r_3 s_2 t_2 + q_1 r_3 s_2 t_3 
+ q_1 r_3 s_2 t_4 + q_1 r_3 s_3 t_1 + q_1 r_3 s_3 t_2 + q_1 r_3 s_3 t_3 \nonumber\\
&+& 
q_1 r_3 s_3 t_4 + q_1 r_3 s_4 t_1 + q_1 r_3 s_4 t_2 + q_1 r_3 s_4 t_3 + q_1 r_3 s_4 t_4 + q_2 r_1 s_1 t_1 
+ q_2 r_1 s_2 t_1 + q_2 r_1 s_2 t_2 + q_2 r_1 s_2 t_3 + q_2 r_1 s_2 t_4 \nonumber\\
&+& 
q_2 r_2 s_1 t_1 + q_2 r_2 s_2 t_1 + q_2 r_2 s_2 t_2 + q_2 r_2 s_2 t_4 + q_2 r_2 s_4 t_1 + q_2 r_3 s_1 t_1 
+ q_2 r_3 s_1 t_3 + q_2 r_3 s_2 t_1 + q_2 r_3 s_2 t_2 + q_2 r_3 s_2 t_3 \nonumber\\
&+& 
q_2 r_3 s_2 t_4 + q_2 r_3 s_4 t_1 + q_2 r_3 s_4 t_2 + q_2 r_3 s_4 t_3 + q_2 r_3 s_4 t_4 + q_2 r_4 s_1 t_1 
+ q_2 r_4 s_2 t_1 + q_2 r_4 s_2 t_2 + q_2 r_4 s_2 t_3 + q_2 r_4 s_2 t_4 \nonumber\\
&+& 
q_3 r_1 s_1 t_1 + q_3 r_1 s_2 t_1 + q_3 r_1 s_2 t_3 + q_3 r_1 s_2 t_4 + q_3 r_1 s_3 t_1 + q_3 r_2 s_1 t_1 
+ q_3 r_2 s_2 t_1 + q_3 r_2 s_2 t_4 + q_3 r_2 s_3 t_1 + q_3 r_2 s_4 t_1\nonumber\\ 
&+& 
q_3 r_3 s_1 t_1 + q_3 r_3 s_1 t_3 + q_3 r_3 s_2 t_1 + q_3 r_3 s_2 t_3 + q_3 r_3 s_2 t_4 + q_3 r_3 s_3 t_1 
+ q_3 r_3 s_3 t_3 + q_3 r_3 s_4 t_1 + q_3 r_3 s_4 t_3 + q_3 r_3 s_4 t_4 \nonumber\\
&+& 
q_3 r_4 s_1 t_1 + q_3 r_4 s_2 t_1 + q_3 r_4 s_2 t_4 + q_3 r_4 s_3 t_1 + q_3 r_4 s_4 t_1 + q_4 r_1 s_1 t_1 
+ q_4 r_1 s_1 t_2 + q_4 r_1 s_2 t_1 + q_4 r_1 s_2 t_2 + q_4 r_1 s_2 t_3 \nonumber\\
&+& 
q_4 r_1 s_2 t_4 + q_4 r_1 s_3 t_1 + q_4 r_1 s_3 t_2 + q_4 r_1 s_3 t_3 + q_4 r_1 s_3 t_4 + q_4 r_2 s_1 t_1 
+ q_4 r_2 s_1 t_2 + q_4 r_2 s_2 t_1 + q_4 r_2 s_2 t_2 + q_4 r_2 s_2 t_4 \nonumber\\
&+& 
q_4 r_2 s_3 t_1 + q_4 r_2 s_3 t_2 + q_4 r_2 s_3 t_4 + q_4 r_2 s_4 t_1 + q_4 r_2 s_4 t_2 + q_4 r_3 s_1 t_1 
+ q_4 r_3 s_1 t_2 + q_4 r_3 s_1 t_3 + q_4 r_3 s_2 t_1 + q_4 r_3 s_2 t_2 \nonumber\\
&+& 
q_4 r_3 s_2 t_3 + q_4 r_3 s_2 t_4 + q_4 r_3 s_3 t_1 + q_4 r_3 s_3 t_2 + q_4 r_3 s_3 t_3 + q_4 r_3 s_3 t_4 
+ q_4 r_3 s_4 t_1 + q_4 r_3 s_4 t_2 + q_4 r_3 s_4 t_3 + q_4 r_3 s_4 t_4 \nonumber\\
&+& 
q_4 r_4 s_1 t_1 + q_4 r_4 s_1 t_2 + q_4 r_4 s_1 t_3 + q_4 r_4 s_2 t_1 + q_4 r_4 s_2 t_2 + q_4 r_4 s_2 t_3 
+ q_4 r_4 s_2 t_4 + q_4 r_4 s_3 t_1 + q_4 r_4 s_3 t_2 + q_4 r_4 s_3 t_3 \nonumber\\
&+& q_4 r_4 s_3 t_4 + q_4 r_4 s_4 t_1 + q_4 r_4 s_4 t_2 + q_4 r_4 s_4 t_3 + q_4 r_4 s_4 t_4.
\nonumber
\end{eqnarray}}
\vspace{-40pt}
\begin{equation}
\end{equation}
The remaining four $w_i$ can be obtained via the interchange symmetry (\ref{E:S5}).

\enlargethispage{20pt}
Then, as long as $w_1+w_2+w_3+w_4+w_5$ is nonzero, the row vector
\begin{equation}
\pi_5 = {[w_1,w_2,w_3,w_4,w_5]\over w_1+w_2+w_3+w_4+ w_5}
\end{equation}
is a properly normalized (row) probability vector.

 It is easy (if now considerably more than somewhat tedious) to check that it is the equilibrium probability vector
\begin{equation}
\label{E:pi5}
\pi_5 \; P_5 = \pi_5. 
\end{equation}
These explicit formulae for $5\times 5$ matrices are sufficiently complex to be most useful when viewed as a consistency check on the general analysis developed above.
\end{itemize}

\bigskip
\hrule\hrule\hrule

\section{Random walks on graphs}
A particularly common example of a stochastic process is the act of taking a random walk on a graph.
Let $A_n =[a_{ij}]$ be the adjacency matrix counting the number of links from node $i$ to node $j$, 
which we will take to be a non-negative integer. Then
\begin{equation}
\d_n = A_n u_n
\end{equation}
counts the ``out degree'' from each node. The vector $\d_n$ maps into a diagonal matrix in a standard way:
\begin{equation}
\d_n \to D_n: \qquad [d_1,\dots d_n] \to 
\left[\begin{array}{cccc}
d_1& 0 & \cdots &0\\
0& d_2 & \cdots &0\\
\vdots &\vdots &\ddots & \vdots\\
0 & 0 & \cdots & d_n
\end{array}\right].
\end{equation}
Then $P_n = D_n^{-1} \; A_n$ is a properly normalised non-negative stochastic matrix defined on the graph.
Explicitly $P_n u_n = u_n$.

Instead of working with the singular matrix $I_n-P_n$, let us now work with the singular matrix $D_n-A_n$. 
As long as the singular (zero) eigenvalue of $D_n-A_n$ has multiplicity 1, it follows that $\rk(D_n-A_n)=n-1$ and so the adjugate matrix 
$\adj(D_n-A_n)$ has rank 1. Thence
$\adj(D_n-A_n)=  u_n \omega_n $ with the row vector $\omega_n$ still to be determined.

But by our previous arguments we see that $(\omega_n)_i = M_{ii}(D_n-A_n)$. Note that $D_n-A_n$ is a matrix of integers, so the minors will all be integers. Note that $D_n-A_n$, and its principal sub-matrices, are all Z-matrices, so the principal minors are all non-negative integers~\cite{Topics}. 

Finally note $\omega_m (D_n-A_n)=0 = (\omega_m D_n)(I_n - P_n)$ so $ \omega_m D_n\propto \pi_n$.
Thence we see that $(\pi_n)_i \propto d_i M_{ii}(D_n-A_n)$, and normalizing this we have
\begin{equation}
\pi_n = {[ d_1M_{11}(D_n-A_n), \cdots, d_i M_{ii}(D_n-A_n)\cdots , d_n M_{nn}(D_n-A_n)]
\over \sum_i d_i M_{ii}(D_n-A_n)}.
\end{equation}
The advantage of doing things this way is that now all the terms in the numerator are integers, as are all the terms in the denominator. The only non-integer part of the computation is the division carried out as the very last step.

\bigskip
\hrule\hrule\hrule
\section{Discussion}
Overall, I feel that the results presented herein are scientifically interesting for two very distinct reasons:
\begin{itemize}
\item Pedagogically the explicit stochastic equilibrium vector for $3\times 3$ stochastic matrices is simple enough to be used as an undergraduate  teaching tool. (The $4\times 4$ case is perhaps best kept in reserve for advanced undergraduates or beginning graduate students.)
\item The $n\times n$ case provides one with some decidedly non-trivial analytic results, and uses some not entirely trivial matrix analysis. Ultimately, for irreducible stochastic matrices the computation is reduced to finding the $n$ principal minors of the singular matrix $I_n-P_n$. For reducible stochastic matrices, focus on the irreducible sub-matrices.
There appears to be some hope for further extending these results in various ways. 
\end{itemize}


\bigskip
\hrule\hrule\hrule
\bigskip
\clearpage

\end{document}